# The screening effects influence on the nano-domain tailoring in ferroelectrics-semiconductors

**Anna N. Morozovska**[*,1], and **Eugene A. Eliseev**[**,2]

[1]  V.Lashkaryov Institute of Semiconductor Physics, National Academy of Science of Ukraine, 41, pr. Nauki, 03028 Kiev, Ukraine

[2]  Institute for Problems of Materials Science, National Academy of Science of Ukraine, 3, Krjijanovskogo, 03142 Kiev, Ukraine

We calculate the realistic sizes of nano-domains recorded by the electric field of atomic force microscope tip in $BaTiO_3$ and $LiNbO_3$ ferroelectric-semiconductors in contrast to the over-estimated ones obtained in the previous works. We modified the existing models of domain formation allowing for the Debye screening, recharging of sluggish surface screening layers caused by emission current between the tip apex and the domain butt surface and the redistribution of domain depolarization field induced by the charged tip apex. We have shown that the depolarization field energy of the domain butt, Debye screening effects and field emission at high voltages lead to the essential decrease of the equilibrium domain sizes. We obtained, that the domain length and radius do not decrease continuously with voltage decrease: the domain appears with non-zero length and radius at definite critical voltage. Such "threshold" domain formation is similar to the first order phase transition and correlates with recent theoretical and experimental investigations.

## 1   Introduction

Submicron spatial regions with reversed spontaneous polarization called micro- and nano- ferroelectric domains have been tailored in many ferroelectrics [1]-[5]. It is clear from general point of view that domains formation can be caused by strong local electric fields with definite polarity. Recently one and two dimensional arrays of spike-like nano-domains have been fabricated in $LiNbO_3$ [1], $LiTaO_3$ [2], $Pb(Zr,Ti)O_3$ [3], $BaTiO_3$ [4], $RbTiOPO_4$ and $RbTiOAsO_4$ [5] ferroelectric crystals with the help of electric fields caused by atomic force microscope (AFM) tip. Obtained nano-domain arrays could be successfully used in modern large-capacity memory devices and light converters based on second harmonic generation. So the possibilities of information recording in the ferroelectric media have been open, if only the optimization problem of high-speed writing nano-domains with high density, stability and fully controllable reversibility would be solved. First of all it is necessary to record the stable domain "dots" with minimum width in the appropriate ferroelectric medium. To realize this idea, one has to determine the dependences of domain radius and length on voltage applied to the AFM tip and ferroelectric medium characteristics either empirically or theoretically. To our mind for the correct description of the numerous experimental results simple modelling seems rather urgent, but present phenomenological models give incomplete description of the nano-domain tailoring owing to the following reasons.

[*] Corresponding author: e-mail: morozo@i.com.ua  
[**] e-mail: eliseev@i.com.ua

- The phenomenological description of the nucleation processes in the perfect dielectric during ferroelectric polarization switching proposed by Landauer [6] should be applied to the domain formation with great care, because in this model the depolarization field is partially screened by the free charges on the metallic electrodes. When modelling the microdomains formation such upper electrode will completely screen the interior of ferroelectric from the AFM tip electric field, thus no external source would induce the polarization reversal. Only homogeneous external field can be applied to such polar dielectric covered with metallic electrodes.
- Theoretical modelling of equilibrium ferroelectric domains recorded by AFM tip proposed in the paper [7] considers tip electric field inside the perfect dielectric-ferroelectric with free surface, i.e. without any screening layer or upper electrode, but the semi-ellipsoidal domain depolarization field was calculated in Landauer model as if the perfect external screening expected. As a result they obtained significantly over-estimated values of domain radius at high voltages [1].
- In our recent papers [8], [9] we try to overcome the aforementioned discrepancies, taking into consideration screening layers of immovable surface charges and semiconductor properties revealed by the most of ferroelectrics [10], [11]. Really, at distances $\Delta R$ between the tip apex and sample surface more then several nm and relatively low applied voltages this screening layer maintains its negative charge during the domain formation owing to the traps sluggishness. However, in the most of experiments [1]-[5] $\Delta R < 1nm$ and thus recharging of this layer is quite possible due to the field emission caused by the strong attraction of trapped carriers by the positively charged AFM tip. Therefore recharging of surface traps caused by the emission current should be taken into account at least at high voltages.
- If AFM tip does not touch the surface, its electric potential distribution inside the crystal could be analyzed using the effective point charge model [6], [7], [12] that ignores electroelastic coupling in the ferroelectric. However effective point charge did not influence at all on the depolarization field created by the domain. But it is obvious, that if the distance between the tip apex and the sample surface is much less than the tip radius of curvature [12], depolarization field causes noticeable free charges redistribution on the metallic tip apex and the depolarization field energy decreases. However these effects were not taken into account in the previous works [7]-[9].

In the present paper for the first time we consider the influence of all aforementioned effects: depolarization field created by the domain butt, Debye screening effects, field emission at high voltages and the depolarization field redistribution induced by the charged tip apex. We have shown that depolarization field of the domain butt, Debye screening and emission current lead to the essential decrease of the equilibrium domain sizes. As a result we obtained the realistic values of domain radius and length recorded in BaTiO$_3$ and LiNbO$_3$ crystals in a wide range of applied voltages.

## 2　Phenomenological description

For description of the charged tip electric field we use the spherical model, in which the tip is represented by the charged sphere with radius $R_0$ located at the distance $\Delta R$ from the sample surface. The voltage $U$ is applied between the tip and ground electrode. The validity of such assumptions for dielectric sample and more sophisticated models are discussed in [12].

Hereinafter we use the model of the rigid ferroelectric with dielectric permittivity ε, displacement $\mathbf{D} = \varepsilon \cdot \mathbf{E} + 4\pi \mathbf{P}_S$ and electric field $\mathbf{E} = -\nabla \varphi(\mathbf{r})$. We choose constant spontaneous polarization $+P_S$ inside and $-P_S$ outside the semi-ellipsoidal domain (see Fig.1).

In ferroelectric-semiconductor the Schottky barriers, band bending, field effects as well as Debye screening cause surface charge layer that effectively shields the interior of the sample from the strong homogeneous depolarization field $E_d = -4\pi P_S$ [11]. Surface charges $\sigma_S = -P_S$ are captured on the sluggish trap levels before the domain formation [13]. These surface charges are almost immovable during the polarization reversal at low tip electric field $E_t \sim U/\Delta R$. Usually $\Delta R < 1nm$ [1]-[5], thus the field emission is quite possible at high voltages $U \geq U_m$. Keeping in mind approach proposed in [14] for current-voltage characteristics of ferroelectric tunnel junctions and interface screening model evolved in [15], we assume that the emission current $J_e \sim \exp(-U_m/U)$. At high voltages the amount of emitted carriers are quite enough to completely screen the reversed polarization of the domain butt-end, i.e. $\sigma_S \to +P_S$ at $U \gg U_m$. So, the equilibrium surface charge density $\sigma_S$ has the form:

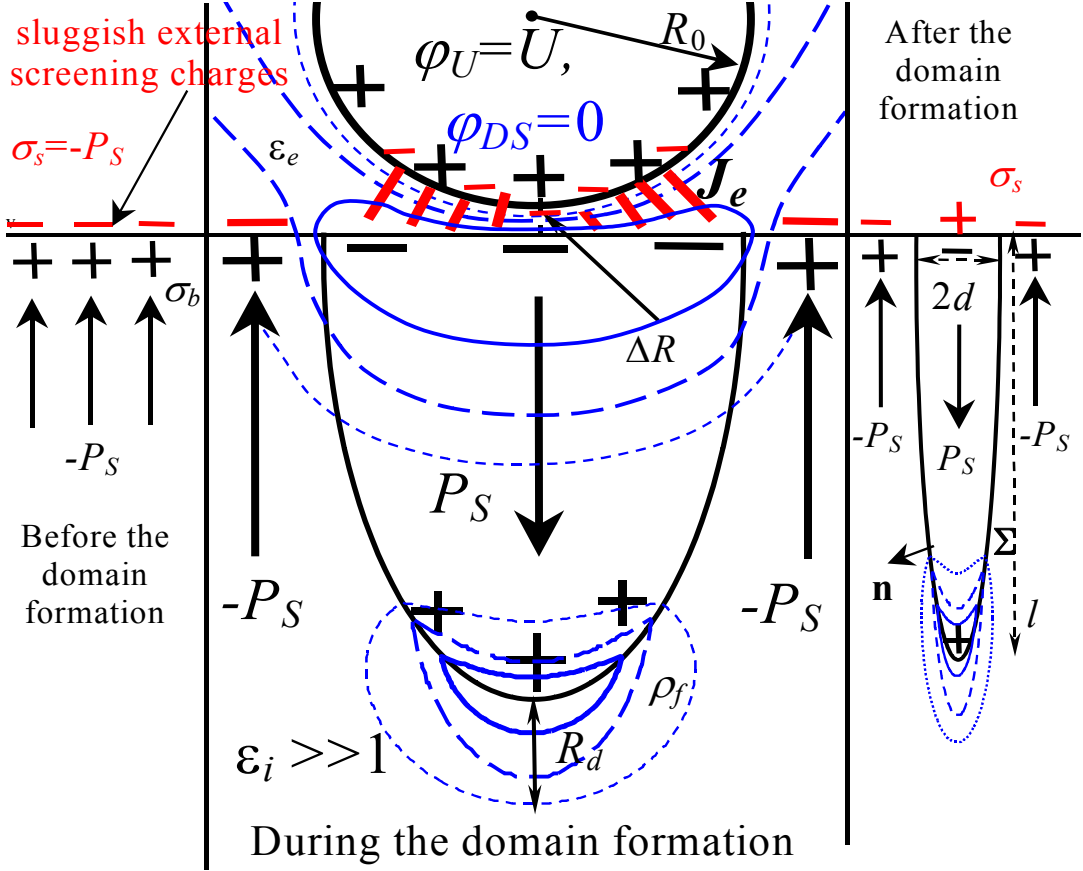

**Fig. 1**  Domain formation induced by positively charged AFM tip. $U$ is applied voltage, $\Delta R$ is the distance between the tip apex and the sample surface, $R_0$ is tip radius of curvature, $d$ is semi-ellipsoid radius, $l$ is semi-ellipsoid major axis, $R_d$ is Debye screening radius, $P_S$ is spontaneous polarization, $\sigma_S$ is surface charges captured on the trap levels, $\sigma_b$ is bound charges related to $P_S$ discontinuity, $J_e$ is carriers emission current, $\rho_f$ is free charge density. Curves are isopotential lines of depolarization field caused by semi-ellipsoid domain.

$$\sigma_S = \begin{cases} -P_S + 2P_S \exp(-U_m/U) & \sqrt{x^2+y^2} < d \\ -P_S & \sqrt{x^2+y^2} > d \end{cases} \quad (1)$$

The carriers' emission characteristic voltage $U_m$ complexly depends over the distance $\Delta R$, dielectric permittivity $\varepsilon$, Debye screening radius $R_d$ and other sample-tip material parameters.

Another important experimental fact should be taken into consideration for correct theoretical description of nanodomains tailoring using AFM tip [16], [17]. Molotskii [16] assumes, that a water meniscus appears between the AFM tip apex and a sample surface due to the air humidity. For instance, scanning tunnelling microscopy measurements on Ti surface show that within the interval 20%–50% humidity water layer thickness increases linearly from 50 to 100 nm (see [17] and ref. therein). The authors argue that such strong water adsorption is based on an oxide layer on the Ti surface. Moreover, if AFM tip is wettable, its apex with curvature 25-50 nm can be completely covered with water [18]. Hereinafter we regard that this region has effective dielectric permittivity $\varepsilon_e$ close to 81.

Inside the semiconductor $|Ze\varphi(\mathbf{r})/k_B T| \ll 1$ and thus the screening of electric field $\mathbf{E} = -\nabla\varphi(\mathbf{r})$ is realized by free charges with bulk density $\rho_f(\mathbf{r}) \approx -\varepsilon_i \varphi(\mathbf{r})/4\pi R_d^2$ and Debye screening radius $R_d$ (see Appendix A). The spatial distribution of the electrostatic potential should be determined from the Maxwell equation $\varepsilon_i \Delta\varphi(\mathbf{r}) = -4\pi\rho_f(\mathbf{r})$ supplemented by the interfacial conditions $D_{n\,\text{int}} = D_{n\,\text{ext}}$ on the semi-ellipsoidal domain surface $\Sigma$, $D_{n\,\text{ext}} - D_{n\,\text{int}} = 4\pi\sigma_S$ on the free surface $z = 0$ and potential disappearance at the bottom electrode. Usually $\varepsilon_i \gg 1$, $R_d \sim (10^2 - 10^4)\,nm$ [11], [19] and sample thickness $h \sim (10^3 - 10^5)\,nm$, so $\exp(-h/R_d) \ll 1$, and we obtain the following boundary problem:

$$\Delta\varphi_0(\mathbf{r}) = 0, \qquad z \leq 0,$$
$$\varphi_0|_{\mathbf{r}\in tip} = U, \qquad \varphi_0(z=0) = \varphi(z=0),$$
$$\left(\varepsilon_e \frac{\partial \varphi_0}{\partial z} - \varepsilon_i \frac{\partial \varphi}{\partial z}\right)\bigg|_{z=0} = \begin{cases} 4\pi(\sigma_S - P_S), & \sqrt{x^2+y^2} < d \\ 0, & \sqrt{x^2+y^2} > d \end{cases} \qquad (2)$$
$$\Delta\varphi(\mathbf{r}) - \frac{\varphi(\mathbf{r})}{R_d^2} = 0, \qquad z \geq 0,$$
$$\varepsilon_i \left(\frac{\partial \varphi_{\text{int}}}{\partial n} - \frac{\partial \varphi_{\text{ext}}}{\partial n}\right)\bigg|_{\Sigma} = 8\pi(\mathbf{P}_S \mathbf{n})|_{\Sigma}, \qquad \varphi(z=h) = 0$$

In order to apply all the following results to the anisotropic semiconductor one can make the substitution: $z \to z\sqrt{\varepsilon_a/\varepsilon_c}$ (e.g. $l \to l\sqrt{\varepsilon_a/\varepsilon_c}$), $\varepsilon_i \to \sqrt{\varepsilon_c\varepsilon_a}$, $R_d^2 = \varepsilon_a k_B T/4\pi e^2 n_d$. Here $\varepsilon_a$ and $\varepsilon_c$ are anisotropic dielectric permittivity values perpendicular and along the polar axis $z$.

The solution of (2) can be found by means of the integral transformations, namely we obtained the potential in the form $\varphi(\mathbf{r}) = \varphi_U(\mathbf{r}) + \varphi_D(\mathbf{r})$, where:

**a)** The electric field potential $\varphi_U(\mathbf{r})$ is created by the positively charged AFM tip with radius $R_0$ and centre located in the point $(0, 0, -z_0)$, hereinafter $z_0 = R_0 + \Delta R$. $\varphi_U(\mathbf{r})$ is calculated in Appendix A with the help of images method (see (A.5) and [12]). The series for $\varphi_U(\mathbf{r})$ is rather cumbersome, only in the cases $\Delta R \ll R_0$ and $\Delta R \gg R_0$ its interpolation acquires a relatively simple form:

$$\varphi_U(\mathbf{r}) \approx \begin{cases} UR_0 \int_0^\infty dk \dfrac{2k\cdot\exp\!\left(-z\sqrt{k^2+R_d^{-2}}-kz_0\right) J_0\!\left(k\sqrt{x^2+y^2}\right)}{\varepsilon_e k + \varepsilon_i\sqrt{k^2+R_d^{-2}} + \left(\varepsilon_e k - \varepsilon_i\sqrt{k^2+R_d^{-2}}\right)\exp(-2k\Delta R)}, & \Delta R \ll R_0 \\[2ex] UR_0 \int_0^\infty dk \dfrac{2k\cdot\exp\!\left(-z\sqrt{k^2+R_d^{-2}}-kz_0\right) J_0\!\left(k\sqrt{x^2+y^2}\right)}{\varepsilon_e k + \varepsilon_i\sqrt{k^2+R_d^{-2}}}, & \Delta R \gg R_0 \end{cases} \qquad (3)$$

**b)** The depolarization field potential $\varphi_D(\mathbf{r})$ is created by polarization reversal inside the semi-ellipsoidal domain with radius $d$ and length $l$. In the case of prolate semi-ellipsoidal domain with $d \ll l$ the potential $\varphi_D(\mathbf{r}) = \varphi_{DS}(\mathbf{r}) + \varphi_{DE}(\mathbf{r})$ is calculated in Appendix B. The series for the surface potential $\varphi_{DS}(\mathbf{r})$ created by the domain butt is rather cumbersome (see (B.8-9)). Only in particular cases its interpolation acquires relatively simple form, namely:

$$\varphi_{DS}(\mathbf{r}) = 4\pi(\sigma_S - P_S)\int_0^\infty dk \frac{J_0(k\sqrt{x^2+y^2})}{\varepsilon_e k + \varepsilon_i\sqrt{k^2+R_d^{-2}}} \exp(-z\sqrt{k^2+R_d^{-2}})(d\cdot J_1(kd) - 2k\Psi(k))$$

$$\Psi(k) \approx \begin{cases} z_0^2 \dfrac{\exp(-k R_0^2/z_0) - \exp(-k R_0^2\sqrt{1+(d/z_0)^2}/z_0)}{R_0(\varepsilon_e k + \varepsilon_i\sqrt{k^2+R_d^{-2}})} & \text{at } \Delta R \gg R_0 \text{ or } d \geq R_0 \\ \dfrac{d\cdot J_1(kd)\exp(-2k\Delta R)}{\varepsilon_i\sqrt{k^2+R_d^{-2}} + \varepsilon_e k - \exp(-2k\Delta R)(\varepsilon_i\sqrt{k^2+R_d^{-2}} - \varepsilon_e k)} & \text{at } \Delta R \ll R_0 \text{ and } d \ll R_0 \end{cases} \quad (4)$$

The term $\varphi_{DE}(\mathbf{r})$ is the potential created by the polarization reversal inside the domain [9]. In the case when screening radius $R_d$ is larger then curvature $r_C = d^2/l$ of the semi-ellipsoid apex, the potential $\varphi_{DE}(\mathbf{r})$ inside the spike semi-ellipsoidal domain acquires the form:

$$\varphi_{DE}(\mathbf{r}) \approx \frac{8\pi P_S}{\varepsilon_i}\frac{d^2}{l^2}\left(\ln\left(\frac{2l}{d}\right)-1\right)\exp\left(-\frac{l-\sqrt{z^2+l^2(x^2+y^2)/d^2}}{R_d}\right)\cdot z \quad (5)$$

Note, that the surface $l - \sqrt{z^2+l^2(x^2+y^2)/d^2} = R_d$ where depolarization field is mainly concentrated, corresponds to the "screening" ellipsoid $(x^2+y^2)l^2/d^2 + z^2 = (l-R_d)^2$, with the same ellipticity $d/l$ as the domain one and semi-axes $(l-R_d)$, $d\sqrt{1-R_d/l}$. The density of the screening charges $\rho_f \sim \varphi_{DE}(\mathbf{r})$ depends on the curvature of domain surface $\Sigma$. For example, the charge density accommodated near the semi-ellipsoid domain apex $z=l$ (where spontaneous polarization vector is normal to the domain surface) is maximal (see Fig.1). Surface potentials are usually neglected in papers (see e.g. [7]). To our mind, they could not be neglected in comparison with $\varphi_{DE}(\mathbf{r})$ at least near the sample surface, where $\varphi_{DE} = 0$.

The electrostatic energy of ferroelectrics is $\Phi_{el} = \int dv(\mathbf{D}\cdot\mathbf{E} - 4\pi \mathbf{P}_S\cdot\mathbf{E})/8\pi$ (see chapter 2 in [20]). The excess of electrostatic energy $\Delta\Phi_{el}(d,l)$ caused by the origin of the semi-ellipsoidal domain with reversed polarization $-P_S \to +P_S$ is considered in details in Appendix C. Its general expression is:

$$\Delta\Phi_{el}(d,l) \approx \Delta\Phi_\rho(d,l) + \Delta\Phi_\sigma(d,l)$$

$$\Delta\Phi_\rho = -\frac{\varepsilon_i}{8\pi R_d^2}\int_{z>0} dv\left[(\varphi_U+\varphi_D)^2 - \varphi_U^2\right], \quad (6)$$

$$\Delta\Phi_\sigma = \int_{\Sigma(z>0)} ds(\mathbf{P}_S\cdot\mathbf{n})(\varphi_U+\varphi_D) + \int_{\substack{(x^2+y^2)\leq d^2 \\ z=0}} dxdy\,(\varphi_U+\varphi_D)\frac{\sigma_S - P_S}{2}$$

The excess of electrostatic energy $\Delta\Phi_{el}(d,l) = \Phi_U(d,l) + \Phi_D(d,l)$ and domain wall surface energy $\Phi_C(d,l)$ contributes into the thermodynamic potential $\Phi(d,l)$:

$$\Phi(d,l) = \Phi_U(d,l) + \Phi_D(d,l) + \Phi_C(d,l) \quad (7)$$

Below we consider every term in (6) with accuracy $O(d^2/l^2)$ and under the conditions $d \ll l$ and $\Delta R \ll R_0$ typical for the most of experiments [1]-[5].

**a)** Then excess of energy $\Phi_U(d,l)$ is caused by interaction between the AFM tip electric field and reversed polarization inside semi-ellipsoidal domain. It was calculated using the approximation (3) for $\varphi_U$. The Pade approximation for interaction energy $\Phi_U(d,l)$ over variable $R_d$ acquires the following form:

$$\Phi_U(d,l) \approx 4\pi(P_S - \sigma_S) U R_0 \varepsilon_e \frac{\varepsilon_i + \varepsilon_e}{\varepsilon_i - \varepsilon_e} \ln\left(\frac{\varepsilon_i + \varepsilon_e}{2\varepsilon_e}\right) \frac{R_d \left(R_0 - \sqrt{R_0^2 + d^2}\right)}{(\varepsilon_i + \varepsilon_e) R_d + 2\varepsilon_i \sqrt{R_0^2 + d^2}} \tag{8}$$

At $R_d \to \infty$, $\varepsilon_e = 1$ and $\sigma_S \to -P_S$ the energy (8) coincides with the one calculated in [7]. Note, that exact expression for $\Phi_U(d,l)$ valid for arbitrary $\Delta R$ and $R_0$ values is given by (C.3). It takes into account the more realistic tip shape in the framework of spherical model by means of the image charges method. However for the electric field at the distances higher than the tip radius the approximation with one effective charge gives qualitatively correct description [12].

**b)** The depolarization field energy $\Phi_D(d,l)$ is caused by polarization reversal within the semi-ellipsoidal domain. The approximation for depolarization field energy $\Phi_D(d,l)$ acquires simple form only at $d \ll l$, $\Delta R \ll R_0$ and $\varepsilon_i \gg 1$, namely:

$$\begin{aligned}
\Phi_D(d,l) &= \Phi_{DE}(d,l) + \Phi_{DS}(d) \\
\Phi_{DE}(d,l) &= \frac{16\pi^2 P_S^2}{3\varepsilon_i} \frac{d^4 R_d (\ln(2l/d) - 1)}{l R_d + 4d^2 (\ln(2l/d) - 1)/3}, \\
\Phi_{DS}(d) &\approx \begin{cases} \pi^2 (\sigma_S - P_S)^2 d^2 \dfrac{R_d (2\varepsilon_e R_d + \varepsilon_i \Delta R)}{(\varepsilon_e R_d + \varepsilon_i \Delta R)^2} \Delta R & at \quad d \ll R_0 \\ \pi^2 (\sigma_S - P_S)^2 \dfrac{d^3 R_d}{(\varepsilon_i + \varepsilon_e) 3\pi R_d / 16 + \varepsilon_i d} & at \quad d \gtrsim R_0/2, \quad \Delta R \neq 0 \end{cases}
\end{aligned} \tag{9}$$

At $d \gg R_d$ one obtains from (8), that the first term $\Phi_{DE}(d,l) \sim d^4/l$ is interaction energy of the real ($z = l$) and imaginary ($z = -l$) bound charges. The last positive term $\Phi_{DS}(d)$ is the intrinsic electrostatic energy of the system "polarized domain butt-charged tip apex", omitted in [7], [21], [22]. It is obvious that in the case of strong indentation limit when the tip is in the mechanical contact with ferroelectric surface and the domain size is limited by the tip sample contact area [21], one can neglect the term $\Phi_{DS}(d)$, because $\Delta R \equiv 0$.

**c)** The correlation surface energy $\Phi_C(d,l)$ of the semi-ellipsoidal domain with $d \ll l$ has the form:

$$\Phi_C(d,l) = \pi d^2 \psi_S + \pi d l \psi_S \frac{1}{\sqrt{1 - (d/l)^2}} \arcsin\left(\sqrt{1 - (d/l)^2}\right) \approx \frac{\pi^2}{2} \psi_S d l \tag{10}$$

We regard domain walls as infinitely thin, with homogeneous surface energy density $\psi_S$.

For the anisotropic ferroelectric-semiconductor the obtained thermodynamic potential of the prolate semi-ellipsoidal domain with $d \ll l$ acquires the form:

$$\begin{aligned}
\Phi(d,l) &\approx \frac{4\pi(P_S - \sigma_S) U R_0 R_d \left(R_0 - \sqrt{R_0^2 + d^2}\right)}{\left(\sqrt{\varepsilon_a \varepsilon_c} + \varepsilon_e\right) R_d + 2\sqrt{\varepsilon_a \varepsilon_c} \sqrt{R_0^2 + d^2}} \varepsilon_e \frac{\sqrt{\varepsilon_a \varepsilon_c} + \varepsilon_e}{\sqrt{\varepsilon_a \varepsilon_c} - \varepsilon_e} \ln\left(\frac{\sqrt{\varepsilon_a \varepsilon_c} + \varepsilon_e}{2\varepsilon_e}\right) + \frac{\pi^2}{2} \psi_S d l + \\
&+ \frac{\pi^2 (\sigma_S - P_S)^2 d^3 R_d}{3\pi R_d \left(\sqrt{\varepsilon_a \varepsilon_c} + \varepsilon_e\right)/16 + \sqrt{\varepsilon_a \varepsilon_c} d} + \frac{16\pi^2 P_S^2}{\sqrt{\varepsilon_a \varepsilon_c}} \frac{d^4 R_d \left(\ln\left(2\sqrt{\varepsilon_a/\varepsilon_c} \cdot l/d\right) - 1\right)}{3\sqrt{\varepsilon_a/\varepsilon_c} l R_d + 4d^2 \left(\ln\left(2\sqrt{\varepsilon_a/\varepsilon_c} \cdot l/d\right) - 1\right)}
\end{aligned} \tag{11}$$

The equilibrium domain sizes minimize the potential (11) and thus satisfy the system of equations:

$$\begin{cases} \dfrac{\partial \Phi(d,l)}{\partial l} = 0, & \dfrac{\partial \Phi(d,l)}{\partial d} = 0, \\ \dfrac{\partial^2 \Phi(d,l)}{\partial^2 l} > 0, & \dfrac{\partial^2 \Phi(d,l)}{\partial^2 d} > 0. \end{cases} \quad (12)$$

In our model (1) the dependence $\sigma_S(U) = -P_S + 2P_S \exp(-U_m/U)$ is known, thus equations (12) for $d(U)$ and $l(U)$ can be solved, namely the dependence of $d$ and $l$ over applied voltage $U$ can be calculated.

The dependence of the thermodynamic potential (11) on the domain length and radius for different values of the applied voltage is represented in the Figs 2. It is clear that for the small voltages it has no minimum at all. With the voltage increase the absolute minimum $\Phi_{\min}(d,l) < 0$ appears at applied voltage $U \sim 0.5V$. Domain has radius $d_{\min} \sim 1-5 nm$, but its length $l_{\min}$ is compatible or smaller than this value. In this case the general form of depolarization factor should be used in Landauer-like energy (9) [6]. Whereas other terms in the potential (11) are valid for the opposite case $l \gg d$, we cannot say whether domains with $d \geq l$ are stable or not. However one can expect that oblate domain appears at the first stage of its growth, moreover the obtained value of $d_{\min}$ is in a reasonable agreement with the one calculated by Kalinin *et. al* [21] for the domain reversal in the strong indentation regime. With the further increase of applied voltage the domain length rapidly increases and the stable prolate domain appears.

So we can conclude that thermodynamic potential (11) has the absolute minimum $\Phi_{\min}(d,l) < 0$ at $\{l = l_{\min}, d = d_{\min}\}$ only when $U > U_{cr}$. At lower voltages $U < U_{cr}$ the domain formation becomes energetically impossible. The value $U_{cr}$ determines the point where the homogeneous polarization distribution becomes absolutely unstable.

Such "threshold" domain formation is similar to the well-known first order phase transition. This result seems quite reasonable, because usually stable ferroelectric nanodomains in thin films can be recorded only above some critical voltage 3-6V (see e.g. [2], [3]), on the other hand threshold voltage was calculated in other models (see e.g. [16], [23]). It seems possible, that the threshold appears when the expanding pressure, acting on the domain wall through the field of the tip, overcomes the compressing pressure [23], i.e. $\partial \Phi(d,l)/\partial d > 0$. Note that in the equilibrium resultant pressure is absent (see (12)), and thus for stable domain formation both approaches are equivalent. However in [23] the compressing pressure is caused by the Landauer depolarization field only, but the latter should also include the domain butt depolarization energy, screening contributions *etc*.

In the next section we present the dependences $d(U)$ and $l(U)$, then compare our results with experiments.

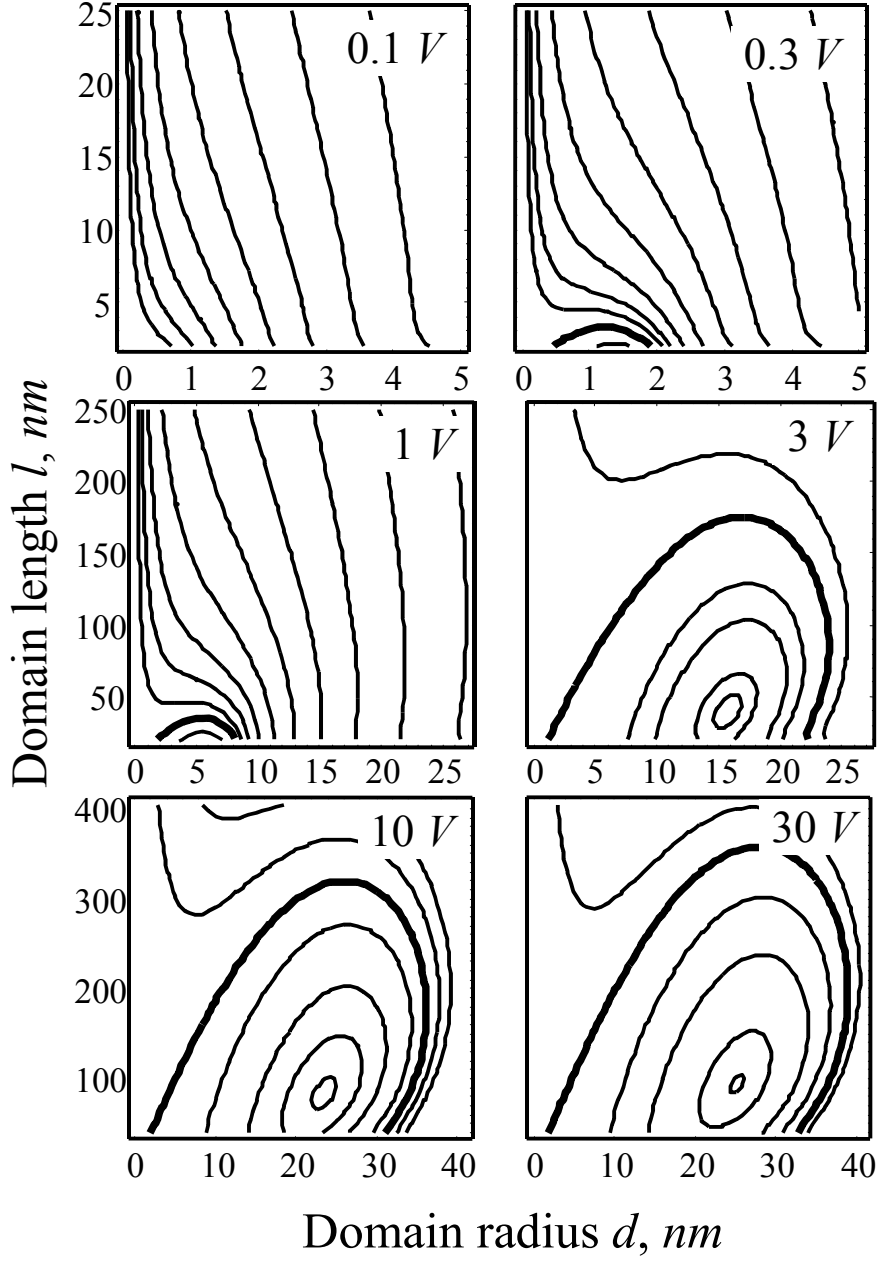

**Fig. 2** Contour lines of the free energy $\Phi(d,l)$ at different voltages $U$ applied to the tip. The bold line marks the contour $\Phi(d,l)=0$. The applied voltage is 0.3 V (a), 0.6 V (b) and 80 V (c). We used typical for BaTiO$_3$ parameters $\psi_S \approx 8\,mJ/m^2$ [24], $P_S \approx 26\,\mu C/cm^2$, $\varepsilon_a = 2000$, $\varepsilon_c = 120$ [25], $R_d = 250\,nm$ and tip radius $R_0 = 25\,nm$, $\Delta R \leq 1\,nm$, carriers' emission characteristic voltage $U_m = 1.25\,V$, $U_{cr} \sim 1\,V$.

## 3  Equilibrium domain sizes: calculations and comparison with experiments

The dependences (12) of equilibrium domain radius $d$ and length $l$ over applied voltage $U$ for different Debye screening radius are shown in the Figs 3. The chosen parameters are typical for BaTiO$_3$ crystals, where nanodomain formation was observed. For example Gruverman *et al*. [26] shown that nanodomains in BaTiO$_3$ can be recorded under the voltage 10V applied to the tip. Eng et al. [4] recorded nanodomains with radius 30nm in BaTiO$_3$ under the voltage 80V applied to the tip.

We would like to underline, that Debye screening not only decreases depolarization field inside the domain, but also it shields the AFM tip electric field inside the sample. As a result, Debye screening radius decrease leads to the decrease of the equilibrium domain sizes (see the lowest curves in the Figs 3). However the critical voltage $U_{cr}$ are almost independent over $R_d$ value at $R_d \geq d$.

Note, that emitted carriers fully compensate not only the depolarization field caused by the reversed polarization of the domain butt (see eqs. (4), (9)), but they simultaneously screen the charged tip electric field, which is the reason of the domain formation. In general case carriers emission leads to the essential decrease of the of the domain sizes at high voltages, namely the domain growth stops at $U >> U_m$.

The aforementioned depolarization field energy (9) of the domain butt essentially decreases the domain sizes even at low voltages [8]. It did not appear in the system considered in [7] due to the complete screening of surface bond charges by the free charge inside the upper electrode (see dotted curves).

Now let us apply our theoretical results to the micro-domain formation in LiNbO$_3$ single crystals using high-voltage AFM. In experiments [1], [22] AFM tip radius was $R_0 = 50 nm$, distance $\Delta R \sim 0.1 nm$, maximum applied voltage pulse value $U_{max} = 4kV$ with duration up to 5min., sample thickness $h = 1 mm$. For LiNbO$_3$ at room temperature $\varepsilon_a = 84$, $\varepsilon_c = 30$, $\psi_S \approx (4-6) mJ/m^2$, $P_S \approx (50-70) \mu C/cm^2$. The comparison of experimental results and our calculations is presented in the Fig.4. The obtained fitting value of Debye screening radius $R_d > 100 \mu m$ is in reasonable agreement with the estimations $R_d \sim (10^4 - 10^6) nm$ valid for the most ferroelectrics-semiconductors with unavoidable growth defects [19].

Let us underline, that the nucleus sizes cannot be smaller than several correlation lengths: for the smaller nuclei the rigid model with $P_S \approx const$ and $\psi_S \approx const$ are invalid, because correlation and size effects begin to play the crucial role [25]. The correlation length cannot be smaller than the several lattice constants, i.e. for LiNbO$_3$ we obtained $d_{min} \geq 1.1 nm$. Keeping in mind this limitation, we obtained rather low value of critical voltage about 1 *V* that corresponds to the atomic scale domains with radius 1.4 *nm* and length 18 *nm*. These values are close to ones obtained by Kalinin *et al.* [21] for the domain tailoring in BaTiO$_3$ crystals.

It should be noted that the results of our work hardly can be compared with experimental data obtained by Terabe *et al*. [27] since in this work the domains with radius ranging from 0.5 to 3.5 μm were reversed in the stochiometric LiNbO$_3$ single crystal plate with thickness 5 μm using the voltage of 40 V. Simple estimation showed that in this system equilibrium domain length considerably exceeds the plate thickness. In this case above-mentioned theory cannot be applied since we neglected the influence of the bottom electrode, however recently we have modified our approach for the nanodomain tailoring in thin ferroelectric films [28].

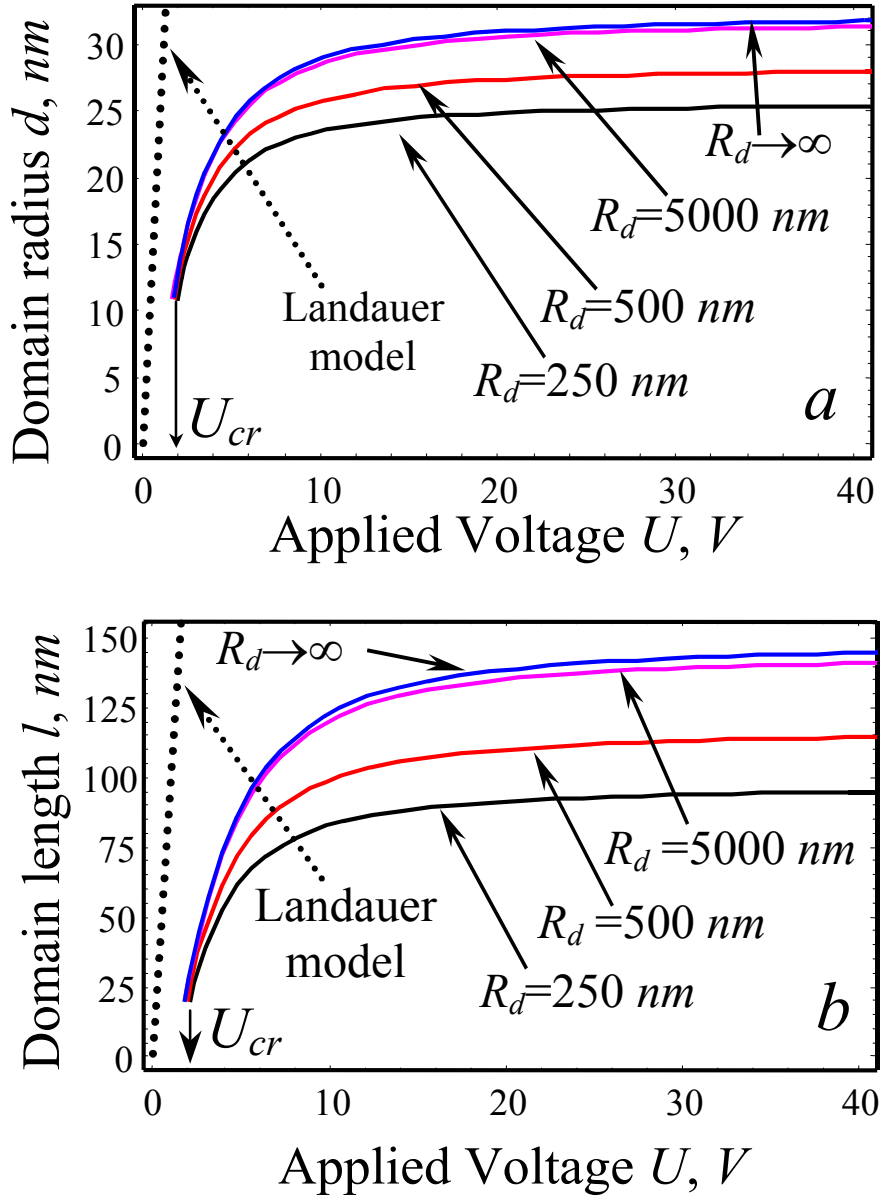

**Figure 3.** Equilibrium domain radius $d$ (a) and length $l$ (b) dependences over applied voltage $U$ for different Debye screening radius $R_d$ (solid curves). Dashed curves represent calculations without domain butt depolarization energy and screening effects (Landauer model [6]). We used typical for BaTiO$_3$ parameters $P_S \approx 26\,\mu C/cm^2$, $\varepsilon_a = 2000$, $\varepsilon_c = 120$ [25], $\psi_S \approx 8\,mJ/m^2$ [24], $\varepsilon_e = 81$ and $\Delta R \le 1\,nm$, $R_0 = 25\,nm$, carriers' emission characteristic voltage $U_m = 1.25\,V$.

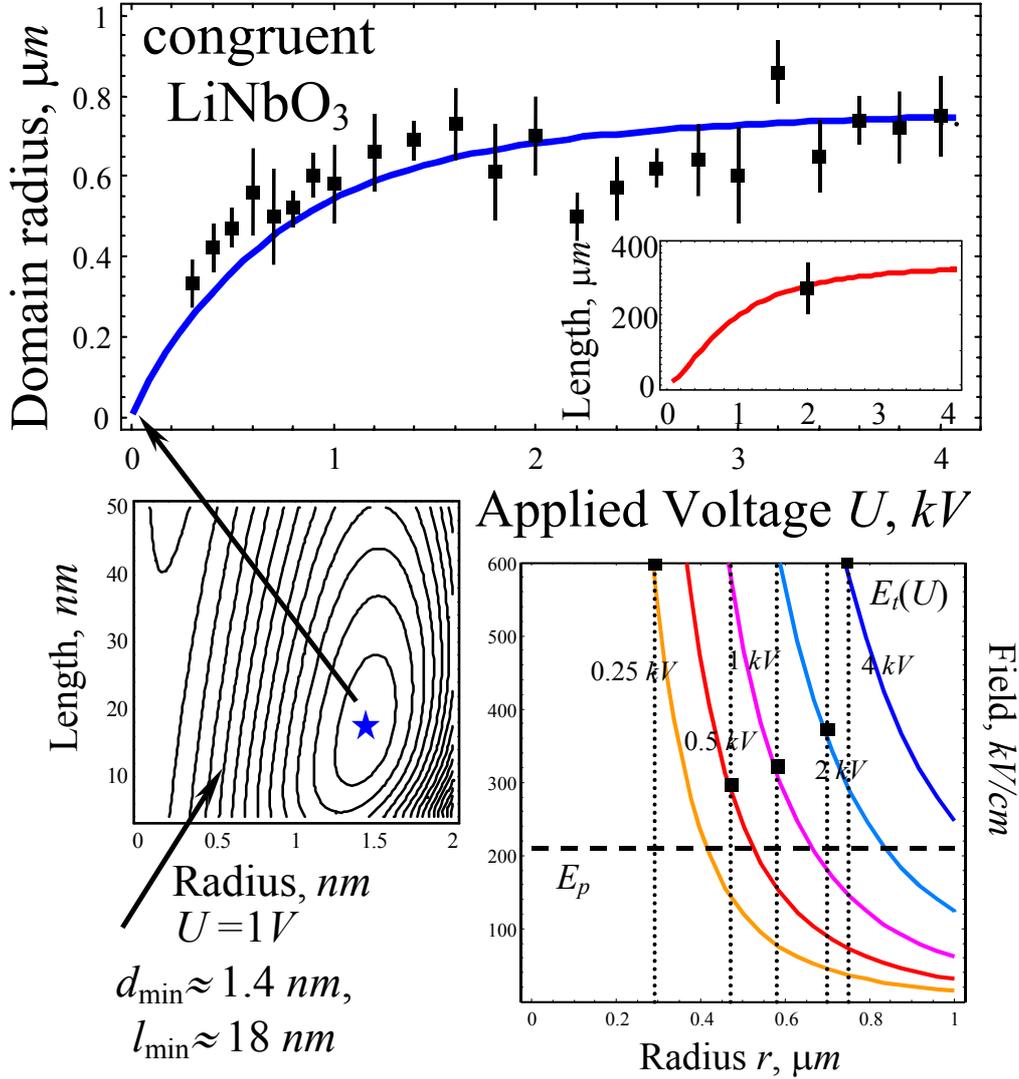

**Figure 4.** Equilibrium domain radius and length over applied voltage $U$ in LiNbO$_3$. Squares with error bars are experimental data from [1] for domain radius $d$, domain length was estimated as 150-250$\mu m$ [22], $R_0 = 50\,nm$, $\Delta R < 1\,nm$. Solid curve is our fitting at $R_d = 500\,\mu m$, $\psi_S = 5\,mJ/m^2$ [29], carriers' emission characteristic voltage $U_m \approx 0.15 \cdot kV$, $P_S \approx 50\,\mu C/cm^2$, $\varepsilon_a = 85$, $\varepsilon_c = 30$, $\varepsilon_e = 81$.

## 4  Discussion

In our consideration we do not take into account the role of pinning centres, although experiment [1] was performed on congruent LiNbO$_3$ samples, which contain numerous imperfections. Rosenman *et al.* [5] pointed out that defects are energy barriers for moving domain walls, which they cannot overcome and estimated the lateral dimension of domains as the region where external electric field exceeds so called pinning field. Sometimes this field determines the experimentally observed coercive field due the pinning-depinning mechanisms of domain walls transition [30], [31]. However the model [5] gives only the incomplite picture of domain formation since it does not take into account the depolarization field and domain wall energy. Furthermore the model [5] does not reveal the saturation of domains radius with applied voltage increase since the higher voltage the wider the region where electric field of the AFM tip exceeds pinning field. In our model we obtain saturation of dependence $d(U)$ allowing for the field emission only. In order to answer the question, which mechanism could be more significant for sub-microdomain recording in LiNbO$_3$, one has to compare the pinning field $E_p$ [5] with the dragging field of the tip $E_t(U)$ at the domain boundary $r = d$. It is appeared experimentally measured value $E_p \approx 210 kV/cm$ [5] is smaller than the dragging field $E_t(U)$ in the region of applied voltages (see squares in the right inset in the Fig.4). The pinning of the domain wall [30] may be important in general case, but the domain kinetics is unfortunately out of the scope of the paper.

Let us discuss the question of domain stability when applied voltage is turned off. In the majority of experiments [1], [2], [22] reversed spike-like domains remain their initial shape and sizes during many days and weeks. This fact is extremely useful for the applications and has the following explanation within the framework of the proposed model. When voltage $U$ is turned off the external field disappears as proportional to $U$, the domain butt depolarization field vanishes due to the final recharging of the surface traps. For the spike-like domains with $d \ll l$ the remained "bulk" depolarization field $E_{DE}(d,l) \sim (d/l)^2 4\pi P_S/\varepsilon$ tends to reverse the domain, but it is appeared too small in comparison with thermodynamic coercive field $E_c \sim 4\pi P_S/\varepsilon$. Also the stabilization process could be related to the pinning of the domain wall [30].

## Conclusion

- We obtained the realistic sizes of nano-domains recorded by the electric field of atomic force microscope tip in BaTiO$_3$ and LiNbO$_3$ ferroelectric-semiconductors, in contrast to the over-estimated ones calculated in the previous papers [7-9].
- We modified the existing models for semi-ellipsoidal domain formation allowing for Debye screening effects, the depolarization energy of the domain butt, recharging of sluggish surface screening layers, field emission between the tip apex and the sample surface originated at high electric fields. We have shown that all these effects lead to the essential decrease of the equilibrium domain sizes.
- We demonstrated, that the domain length and radius do not decrease continuously with voltage decrease: the domain appears with non-zero length and radius at definite critical voltage. Such "threshold" domain formation is similar to the well-known first order phase transition and correlates with recent experimental [2], [3] and theoretical [16], [22], [23] results.
- We hope, that our results will help one to determine the necessary recording conditions and appropriate ferroelectric medium in order to obtain the stable domains with minimum lateral size in a wide range of applied voltages.

## APPENDIX A

For the semiconductor with donors concentration $n_d$, the free charges bulk density $\rho_f$ is determined via electric field potential $\varphi(\mathbf{r})$, as follows:

$$\rho_f(\mathbf{r}) = e\left( Z n_d \exp\left(-\frac{Ze\varphi(\mathbf{r})}{k_B T}\right) - n_0 \exp\left(\frac{e\varphi(\mathbf{r})}{k_B T}\right)\right) \tag{A.1}$$

Hereinafter we consider the case $|Ze\varphi(\mathbf{r})/k_B T| \ll 1$. Taking into account the electro neutrality condition $Zn_d = n_0$, one can find that

$$\rho_f(\mathbf{r}) \approx -\frac{\varepsilon \varphi(\mathbf{r})}{4\pi R_d^2}, \quad R_d^2 = \frac{\varepsilon k_B T}{4\pi e^2 (Z^2 n_d + n_0)} \tag{A.2}$$

Usually $\varepsilon \gg 1$, $R_d \sim (10^2 - 10^4) nm$ and $h \sim (10^3 - 10^5) nm$, so $\exp(-h/R_d) \ll 1$ and hereinafter we regard the bottom electrode located in the infinity. The external electric field potential $\varphi_U(\mathbf{r})$ is created by the spherical tip with radius $R_0$. Its center is located in air the in the point $r_0 = (0,0,-z_0)$ near the boundary of the isotropic semiconductor with dielectric permittivity $\varepsilon$. It could be found from the boundary problem:

$$\Delta\varphi_{U0}(\mathbf{r}) = 0, \qquad z \le 0,$$

$$\Delta\varphi_{U}(\mathbf{r}) - \frac{\varphi_{U}(\mathbf{r})}{R_d^2} = 0, \qquad z \ge 0,$$

$$\varphi_{U0}(z = 0) = \varphi_{U}(z = 0), \qquad \left(\frac{\partial\varphi_{U0}}{\partial z} - \varepsilon\frac{\partial\varphi_{U}}{\partial z}\right)\bigg|_{z=0} = 0, \qquad (A.3)$$

$$\varphi_{U0}\big|_{\mathbf{r}\in AFM_{tip}} = U, \qquad \varphi_{U}(z \to \infty) = 0.$$

The general solution of (A.3) could be found in the form:

$$\varphi_{U0}(\mathbf{r}) = \int_0^\infty dk J_0\left(k\sqrt{x^2 + y^2}\right)\left(\exp(-k \cdot |z + z_0|)A_0(k) + \exp(-k \cdot (z_0 - z))B_0(k)\right)$$

$$\varphi_{U}(\mathbf{r}) = \int_0^\infty dk J_0\left(k\sqrt{x^2 + y^2}\right)\exp\left(-\sqrt{k^2 + R_d^{-2}} \cdot z\right)C_U(k) \qquad (A.4)$$

Hereinafter $J_0$ is Bessel function of zero order, $z_0 = R_0 + \Delta R$. Functions $A_0$, $B_0$ and $C_U$ can be determined from the boundary conditions. Using the method of images, we obtained that:

$$\varphi_{U0}(\mathbf{r}) = UR_0\int_0^\infty dk J_0\left(k\sqrt{x^2 + y^2}\right)\sum_{m=0}^\infty q_m(k)\left(\begin{array}{c}\exp(-k\cdot|z_0 - r_m + z|) + \\ +\exp(-k\cdot(z_0 - r_m - z))\frac{k - \varepsilon\sqrt{k^2 + R_d^{-2}}}{k + \varepsilon\sqrt{k^2 + R_d^{-2}}}\end{array}\right),$$

$$\varphi_{U}(\mathbf{r}) = UR_0\int_0^\infty dk J_0\left(k\sqrt{x^2 + y^2}\right)\sum_{m=0}^\infty 2k q_m(k)\left(\frac{\exp\left(-z\sqrt{k^2 + R_d^{-2}} - k(z_0 - r_m)\right)}{k + \varepsilon\sqrt{k^2 + R_d^{-2}}}\right), \qquad (A.5)$$

$$q_0 = 1, \quad q_m(k) = \left(\frac{\varepsilon\sqrt{k^2 + R_d^{-2}} - k}{\varepsilon\sqrt{k^2 + R_d^{-2}} + k}\right)\cdot\frac{R_0}{2z_0 - r_{m-1}}q_{m-1}(k) \equiv \left(\frac{\varepsilon\sqrt{k^2 + R_d^{-2}} - k}{\varepsilon\sqrt{k^2 + R_d^{-2}} + k}\right)^m\frac{sh(\theta)}{sh((m+1)\theta)},$$

$$r_0 = 0, \quad r_m = \frac{R_0^2}{2z_0 - r_{m-1}} \equiv R_0\frac{sh(m\theta)}{sh((m+1)\theta)}, \qquad ch(\theta) = \frac{z_0}{R_0}$$

Using Zommerfeld formula

$$\frac{\exp\left(-\sqrt{x^2 + y^2 + z^2}/R_d\right)}{\sqrt{x^2 + y^2 + z^2}} = \int_0^\infty dk \frac{k J_0\left(k\sqrt{x^2 + y^2}\right)}{\sqrt{k^2 + R_d^{-2}}}\exp\left(-\sqrt{k^2 + R_d^{-2}}\cdot|z|\right), \qquad (A.6)$$

at $\varepsilon \gg 1$ and $R_d \to \infty$ one obtains from (A.5) well-known potentials [12]

$$\varphi_{U0}(\mathbf{r}) = UR_0\sum_{m=0}^\infty\left(\frac{q_m}{\sqrt{x^2 + y^2 + (z_0 - r_m + z)^2}} - \frac{\varepsilon - 1}{\varepsilon + 1}\frac{q_m}{\sqrt{x^2 + y^2 + (z_0 - r_m - z)^2}}\right) \qquad (A.7a)$$

Inside the sample:

$$\varphi_U(\mathbf{r}) = \sum_{m=0}^{\infty} \frac{2UR_0 q_m}{(\varepsilon+1)\sqrt{x^2 + y^2 + (z_0 - r_m + z)^2}} \tag{A.7b}$$

In order to apply this result to the anisotropic semiconductor one can use the substitution:

$$z \to z/\sqrt{\varepsilon_c/\varepsilon_a}, \quad \varepsilon \to \sqrt{\varepsilon_c \varepsilon_a}, \quad R_d^2 = \frac{\varepsilon_a k_B T}{4\pi e^2 (Z^2 n_d + n_0)}.$$

Now let us find the interpolation for $\varphi_U(\mathbf{r})$ in the case $z_0 \gg R_0$ (i.e. $\Delta R \gg R_0$). In this case $q_{m+1}(k) \ll q_0(k)$, $r_m \ll R_0$, so one obtains from (A.5) that:

$$\varphi_U(\mathbf{r}) \approx UR_0 \int_0^\infty dk J_0\left(k\sqrt{x^2+y^2}\right) \frac{2k \cdot \exp\left(-z\sqrt{k^2+R_d^{-2}} - k z_0\right)}{\varepsilon\sqrt{k^2+R_d^{-2}} + k} \tag{A.8}$$

Now let us find the interpolation for $\varphi_U(\mathbf{r})$ in the case $\Delta R \ll R_0$ (i.e. $\theta \ll 1$). In this case $\varphi_{U0}(z=-\Delta R) \approx U$ and one could easy calculate directly from (A.5) that:

$$\varphi_U(\mathbf{r}) \approx UR_0 \int_0^\infty dk J_0\left(k\sqrt{x^2+y^2}\right) \frac{2k \cdot \exp\left(-z\sqrt{k^2+R_d^{-2}} - k z_0\right)}{k + \varepsilon\sqrt{k^2+R_d^{-2}} + \left(k - \varepsilon\sqrt{k^2+R_d^{-2}}\right)\exp(-2k\Delta R)} \tag{A.9}$$

In order to apply this result to the system consisting of anisotropic semiconductor and isotropic dielectric with permittivity tensors $\{\varepsilon_a, \varepsilon_a, \varepsilon_c\}$ and $\{\varepsilon_e, \varepsilon_e, \varepsilon_e\}$ respectively, one can use the substitution:

$$z \to z/\sqrt{\varepsilon_c/\varepsilon_a}, \quad \varepsilon \to \sqrt{\varepsilon_c \varepsilon_a}, \quad k \pm \varepsilon\sqrt{k^2+R_d^{-2}} \to \left(\varepsilon_e k \pm \sqrt{\varepsilon_c \varepsilon_a}\sqrt{k^2+R_d^{-2}}\right), \tag{A.10a}$$

$$R_d^2 = \varepsilon_a k_B T / 4\pi e^2 (Z^2 n_d + n_0). \tag{A.10b}$$

### APPENDIX B

The depolarization field potential $\varphi_D(\mathbf{r})$ satisfies the Poisson equation with the interfacial conditions $D_{n\,\text{int}} = D_{n\,\text{ext}}$ on the domain surface $\Sigma$ and $D_{n\,\text{ext}} - D_{n\,\text{int}} = 4\pi\sigma_S$ at the surface $z=0$. It can be written as the boundary problem:

$$\begin{gathered}
\Delta\varphi_{D0}(\mathbf{r}) = 0, \quad z \leq 0, \\
\varphi_{D0}(z=0) = \varphi_D(z=0), \quad \left(\frac{\partial\varphi_{D0}}{\partial z} - \varepsilon\frac{\partial\varphi_D}{\partial z}\right)\bigg|_{z=0} = \begin{cases} 4\pi(\sigma_S - P_S), & \sqrt{x^2+y^2} < d \\ 0, & \sqrt{x^2+y^2} > d \end{cases} \\
\Delta\varphi_D(\mathbf{r}) - \frac{\varphi_D(\mathbf{r})}{R_d^2} = 0, \quad z \geq 0 \\
\varepsilon\left(\frac{\partial\varphi_{D\text{int}}}{\partial n} - \frac{\partial\varphi_{D\text{ext}}}{\partial n}\right)\bigg|_\Sigma = 8\pi(\mathbf{P}_S \mathbf{n})\big|_\Sigma, \quad \varphi_{D0}\big|_{\mathbf{r}\in AFM_{tip}} = 0, \quad \varphi_D(z\to\infty) = 0
\end{gathered} \tag{B.1}$$

Firstly let us calculate the part $\varphi_{DE}(\mathbf{r})$ of potential $\varphi_D(\mathbf{r}) = \varphi_{DE}(\mathbf{r}) + \varphi_{DS}(\mathbf{r})$ created by polarized ellipsoid in the infinite isotropic semiconductor with permittivity $\varepsilon$:

$$\Delta\varphi_{DE}(\mathbf{r}) - \frac{\varphi_{DE}(\mathbf{r})}{R_d^2} = 0,$$

$$\varepsilon\left(\frac{\partial\varphi_{DE\,int}}{\partial n} - \frac{\partial\varphi_{DE\,ext}}{\partial n}\right)\bigg|_{\Sigma} = 8\pi(\mathbf{P}_S \mathbf{n})\big|_{\Sigma}, \quad (B.2)$$

$$\varphi_{DE}(z=0) = 0$$

The solution of (B.2) can be found by means of the Green function method and integral transformations, namely we obtained:

$$\varphi_{DE}(\mathbf{r}) = \int_{\Sigma} ds' \frac{2(\mathbf{P}_S \mathbf{n}(\mathbf{r}'))}{\varepsilon} \cdot \frac{\exp(-|\mathbf{r}-\mathbf{r}'|/R_d)}{4\pi|\mathbf{r}-\mathbf{r}'|} =$$
$$= -\frac{4\pi P_S}{\varepsilon}\int_0^\infty dk J_0\left(k\sqrt{x^2+y^2}\right)\int_0^l dz' \frac{d}{dk} J_0\left(kd\sqrt{1-(z'/l)^2}\right) \times \quad (B.3)$$
$$\times \left(sign(z-z')\exp\left(-\sqrt{k^2+R_d^{-2}}\cdot|z-z'|\right) + sign(z+z')\exp\left(-\sqrt{k^2+R_d^{-2}}\cdot|z+z'|\right)\right)$$

Here we integrate over the whole ellipsoid in order to satisfy the condition $\varphi_{DE}(z=0)=0$ and then to continue $\varphi_{DE}(z \le 0) = 0$. One can see that for perfect dielectric $R_d \to \infty$, and thus the solution $\varphi_{DEP}(\mathbf{r}) = \int_{\Sigma} ds' \frac{2(\mathbf{P}_S \mathbf{n})}{4\pi\varepsilon|\mathbf{r}-\mathbf{r}'|}$ of this problem without screening is the known Coulomb potential created by bound surface charge with density $\sigma_b(\mathbf{r}) = 2(\mathbf{P}_S \mathbf{n}(\mathbf{r}))$ and calculated in [20] with the help of ellipsoidal coordinates. In contrast, for ideal conductor $R_d \to 0$, and the solution $\varphi_D(\mathbf{r}) \to 0$ as it should be expected inside the metal.

Allowing for $\sigma_b(\mathbf{r}) \ge 0$, it is easy to obtain from (B.3) the following estimation for potential $\varphi_{DE}(\mathbf{r})$:

$$0 \le \varphi_{DE}(\mathbf{r}) \le \varphi_{DEP}(\mathbf{r})\exp\left(-\frac{r_{\Sigma}(\mathbf{r})}{R_d}\right). \quad (B.4)$$

Hereinafter $r_{\Sigma}(\mathbf{r})$ is the distance between the point $\mathbf{r}$ and the domain boundary $\Sigma(d,l)$. It is clear from (B.4), that for small enough screening radius $R_d$ potential $\varphi_{DE}(\mathbf{r})$ is concentrated inside the layer $r_{\Sigma}(\mathbf{r}) \le R_d$. Keeping in mind exact expression for $\varphi_{DEP}(\mathbf{r})$ and (B.4), we obtained the relatively simple approximation for $\varphi_{DE}(\mathbf{r})$ for elongated domain with $l >> d$ and $R_d \ge d^2/l$:

$$\varphi_{DE}(\mathbf{r}) \le \begin{cases} \frac{8\pi P_S}{\varepsilon}\frac{d^2}{l^2}\left(arcth\left(\sqrt{1-\frac{d^2}{l^2}}\right)-1\right)\exp\left(-\frac{l-\sqrt{z^2+l^2(x^2+y^2)/d^2}}{R_d}\right)\cdot z, & s \le 0 \\ \frac{8\pi P_S}{\varepsilon}\frac{d^2}{l^2}\left(arcth\left(\sqrt{\frac{l^2-d^2}{s+l^2}}\right)-\frac{l}{\sqrt{s+l^2}}\right)\exp\left(-\frac{\sqrt{z^2+l^2(x^2+y^2)/d^2}-l}{R_d}\right)\cdot z, & s \ge 0 \end{cases} \quad (B.5)$$

Here $s(x,y,z)$ is the one of ellipsoidal coordinates $\frac{x^2+y^2}{d^2+s} + \frac{z^2}{l^2+s} = 1$ (s=0 corresponds to the boundary of domain).

Now let us calculate the surface screening potential $\varphi_{DS}(\mathbf{r})$ created by ferroelectric-semiconductor domain butt.

$$\Delta\varphi_{D0}(\mathbf{r}) = 0, \qquad z \le 0,$$

$$\varphi_{D0}\big|_{\mathbf{r}\in AFM_{tip}} = 0, \qquad \varphi_{D0}(z=0) = \varphi_{DS}(z=0),$$

$$\left(\frac{\partial\varphi_{D0}}{\partial z} - \varepsilon\frac{\partial(\varphi_{DS}+\varphi_{DE})}{\partial z}\right)\bigg|_{z=0} = \begin{cases} 4\pi(\sigma_S - P_S), & \sqrt{x^2+y^2} < d \\ 0, & \sqrt{x^2+y^2} > d \end{cases} \tag{B.6}$$

$$\Delta\varphi_{DS}(\mathbf{r}) - \frac{\varphi_{DS}(\mathbf{r})}{R_d^2} = 0, \qquad 0 \le z$$

The solution of (B.6) can be found using expansions from Appendix A. For the spike-like domains with at $d \ll l$ the derivative $\dfrac{\partial\varphi_{DE}(z=0)}{\partial z} = -\dfrac{8\pi P_S}{\varepsilon}\int_0^\infty dk J_0\!\left(k\sqrt{x^2+y^2}\right)\!\left(\dfrac{d}{dk}J_0(kd) - C_D(k)\right) \approx 0$, because

$$C_D(k) = \sqrt{k^2+R_D^{-2}}\int_0^l dz'\exp\!\left(-\sqrt{k^2+R_D^{-2}}\cdot z'\right)\frac{d}{dk}J_0\!\left(kd\sqrt{1-(z'/l)^2}\right) \approx -d\cdot J_1(kd).$$ Substituted (B.3)

into the boundary conditions in (B.6) it is easy to obtain that:

$$\varphi_{D0}(\mathbf{r}) = 4\pi(\sigma_S - P_S)\int_0^\infty dk\, J_0\!\left(k\sqrt{x^2+y^2}\right)\left(\begin{array}{c}\dfrac{d\cdot J_1(kd)\exp(kz)}{k+\varepsilon\sqrt{k^2+R_d^{-2}}} - \psi(k)\exp(-k|z_0+z|) - \\[6pt] -\psi(k)\exp(k(z-z_0))\dfrac{k-\varepsilon\sqrt{k^2+R_d^{-2}}}{k+\varepsilon\sqrt{k^2+R_d^{-2}}}\end{array}\right) \tag{B.7}$$

$$\varphi_{DS}(\mathbf{r}) = 4\pi(\sigma_S - P_S)\int_0^\infty dk\, J_0\!\left(k\sqrt{x^2+y^2}\right)\exp\!\left(-z\sqrt{k^2+R_d^{-2}}\right)\frac{d\cdot J_1(kd) - 2k\psi(k)\exp(-kz_0)}{k+\varepsilon\sqrt{k^2+R_d^{-2}}}$$

The function $\psi(k)$ should be found from the condition $\varphi_{D0}\big|_{\mathbf{r}\in AFM_{tip}} = 0$. Using the method of images, similarly to (A.5) we obtained that:

$$\varphi_{D0}(\mathbf{r}) = 4\pi(\sigma_S - P_S)\int_0^\infty dk\,\frac{k J_0\!\left(k\sqrt{x^2+y^2}\right)}{k+\varepsilon\sqrt{k^2+R_d^{-2}}}\cdot\left(\begin{array}{c}d\cdot\dfrac{J_1(kd)}{k}\exp(kz) - \\[6pt] -\sum_{m=1}^\infty\left(\Psi_m(k,d,z) + \dfrac{k-\varepsilon\sqrt{k^2+R_d^{-2}}}{k+\varepsilon\sqrt{k^2+R_d^{-2}}}\Psi_m(k,d,-z)\right)\end{array}\right), \tag{B.8}$$

$$\varphi_{DS}(\mathbf{r}) = 4\pi(\sigma_S - P_S)\int_0^\infty dk\,\frac{k J_0\!\left(k\sqrt{x^2+y^2}\right)}{k+\varepsilon\sqrt{k^2+R_d^{-2}}}\exp\!\left(-z\sqrt{k^2+R_d^{-2}}\right)\left(d\cdot\frac{J_1(kd)}{k} - \sum_{m=1}^\infty\frac{2k\,\Psi_m(k,d,0)}{k+\varepsilon\sqrt{k^2+R_d^{-2}}}\right)$$

Here we used the designations:

$$\Psi_m(k,d,z) = \int_0^d b\,db\, J_0(k\,\rho_m(b)) \cdot M_m(k,b) \exp(-k \cdot |z_0 - Z_m(b) + z|),$$

$$M_1(k,b) = \frac{R_0}{\sqrt{z_0^2 + b^2}}, \quad M_{m+1}(k,b) = \left(\frac{\varepsilon\sqrt{k^2 + R_d^{-2}} - k}{\varepsilon\sqrt{k^2 + R_d^{-2}} + k}\right) \cdot \frac{R_0 M_m(k,b)}{\sqrt{\rho_m^2(b) + (2z_0 - Z_m(b))^2}},$$

$$Z_1(b) = \frac{z_0 R_0^2}{z_0^2 + b^2}, \quad Z_{m+1}(b) = \frac{R_0^2 \cdot (2z_0 - Z_m(b))}{\rho_m^2(b) + (2z_0 - Z_m(b))^2}, \quad (B.9)$$

$$\rho_1(b) = \frac{b R_0^2}{z_0^2 + b^2}, \quad \rho_{m+1}(b) = \frac{R_0^2 \cdot \rho_m(b)}{\rho_m^2(b) + (2z_0 - Z_m(b))^2},$$

Here $z_0 = R_0 + \Delta R$. The series (B.8) quickly converge only at $R_0/\Delta R \to 0$. In the case $\Delta R \gg R_0$ we obtained:

$$\varphi_{DS}(\mathbf{r}) \approx 4\pi(\sigma_S - P_S) \int_0^\infty dk \frac{k J_0\left(k\sqrt{x^2 + y^2}\right)}{k + \varepsilon\sqrt{k^2 + R_d^{-2}}} \exp\left(-z\sqrt{k^2 + R_d^{-2}}\right) \times$$

$$\times \left(\frac{J_1(kd)d}{k} - \frac{2z_0^2/R_0}{\left(k + \varepsilon\sqrt{k^2 + R_d^{-2}}\right)} \left(\exp\left(-k\frac{R_0^2}{z_0}\right) - \exp\left(-k\frac{R_0^2}{z_0}\sqrt{1 + \frac{d^2}{z_0^2}}\right)\right)\right) \quad (B.10)$$

In the opposite case $\Delta R \ll R_0$ and $\varepsilon R_0 \gg d$ using Laplace method in exp in (B.8-9) we put

$$\rho_m \to b, \, Z_1 \to R_0, \, Z_{m+1} \to \frac{R_0^2}{2z_0 - Z_m}, \, M_m(k) = \left(\frac{\varepsilon\sqrt{k^2 + R_d^{-2}} - k}{\varepsilon\sqrt{k^2 + R_d^{-2}} + k}\right)^{m-1} \frac{R_0 M_m(k)}{2z_0 - Z_m} \text{ and derived similarly to}$$

(A.9):

$$\varphi_{DS}(\mathbf{r}) \approx 4\pi(\sigma_S - P_S) \int_0^\infty dk\, J_0\left(k\sqrt{x^2 + y^2}\right) \frac{d J_1(kd)(1 - \exp(-2k\Delta R))\exp\left(-z\sqrt{k^2 + R_d^{-2}}\right)}{k + \varepsilon\sqrt{k^2 + R_d^{-2}} + \left(k - \varepsilon\sqrt{k^2 + R_d^{-2}}\right)\exp(-2k\Delta R)} \quad (B.11)$$

In particular case $R_0 \gg d$ the tip surface could be regarded as the plain one and (B.11) is exact. In order to apply these results to the anisotropic semiconductor, one can use the substitution (A.10).

## APPENDIX C

The electrostatic energy is created by the surface charges $\sigma_b(\mathbf{r})$ located on the domain surfaces $\Sigma$, $\sigma_b$ and $\sigma_S$ at $z = 0$ related to the spontaneous polarization discontinuity, as well as by the bulk charges $\rho_f(\mathbf{r}) = -\varepsilon\varphi(\mathbf{r})/4\pi R_d^2$ inside the sample. It has the form:

$$\Phi_{el}(\varphi_U,\varphi_D) = \frac{1}{8\pi}\int_V dv (\mathbf{D}\cdot\mathbf{E} - 4\pi\mathbf{P}_S\cdot\mathbf{E}) \equiv \Phi_\rho(\varphi_U,\varphi_D) + \Phi_\sigma(\varphi_U,\varphi_D),$$

$$\Phi_\rho(\varphi_U,\varphi_D) = \frac{1}{2}\int_V dv\,\rho_f(\mathbf{r})\varphi(\mathbf{r}) = -\frac{\varepsilon}{8\pi R_d^2}\int_{z>0} dv(\varphi_U(\mathbf{r}) + \varphi_D(\mathbf{r}))^2, \quad\text{(C.1)}$$

$$\Phi_\sigma = \frac{1}{2}\int_S ds\,\sigma_b\,\varphi = \int_\Sigma ds(\mathbf{P}_S\cdot\mathbf{n})(\varphi_U + \varphi_D) + \int_{\substack{(x^2+y^2)\le d^2\\z=0}} dxdy\,\frac{\sigma_S - P_S}{2}(\varphi_U + \varphi_D).$$

Note, that the first term in $\Phi_\sigma$ acquires the form similar to the one considered in [6], [7] in accordance with Gauss theorem. Let us find the excess of electrostatic energy $\Delta\Phi_{el} = \Phi_{el}(\varphi_U,\varphi_D) - \Phi_{el}(\varphi_U,0)$ caused by polarization reversal inside the domain $-P_S \to +P_S$:

$$\Delta\Phi_{el} \approx \Delta\Phi_\rho + \Delta\Phi_\sigma,$$

$$\Delta\Phi_\rho = -\frac{\varepsilon}{8\pi R_d^2}\int_{z>0} dv\left[(\varphi_U(\mathbf{r}) + \varphi_D(\mathbf{r}))^2 - \varphi_U(\mathbf{r})^2\right], \quad\text{(C.2)}$$

$$\Delta\Phi_\sigma = \int_{\Sigma(z>0)} ds(\mathbf{P}_S\cdot\mathbf{n})(\varphi_U + \varphi_D) + \int_{\substack{(x^2+y^2)\le d^2\\z=0}} dxdy\,\frac{\sigma_S - P_S}{2}(\varphi_U + \varphi_D).$$

In the case $d \ll l$ and $\varepsilon \gg 1$, up to the terms proportional to $\varepsilon^{-2}$ the excess of electrostatic energy acquires the form $\Delta\Phi_{el} = \Phi_D(d,l) + \Phi_U(d,l)$.

1) The interaction energy $\Phi_U(d,l)$ between the domain and the AFM tip acquires the following form (see (A.7) and (B.7)):

$$\Phi_U(d,l) = \int_{\Sigma(z>0)} ds(\mathbf{P}_S\cdot\mathbf{n})\varphi_U + \int_{\substack{(x^2+y^2)\le d^2\\z=0}} dxdy\,\varphi_U\,\frac{\sigma_S - P_S}{2} - \frac{\varepsilon}{4\pi R_d^2}\int_{z>0} dv\,\varphi_D\,\varphi_U \approx$$

$$\approx \frac{4\pi(\sigma_S - P_S)}{\varepsilon + 1}\begin{cases} \sum_{m=0}^{\infty} q_m R_d\,\dfrac{\sqrt{(z_0 - r_m)^2 + d^2} - (z_0 - r_m)}{2\sqrt{(z_0 - r_m)^2 + d^2}} + O\!\left(\!\left(\dfrac{d}{l}\right)^2 \exp(-l/R_d)\right), & R_d \to 0 \\[2mm] \sum_{m=0}^{\infty} q_m\left(\sqrt{(z_0 - r_m)^2 + d^2} - (z_0 - r_m)\right) + O\!\left(\!\left(\dfrac{d}{l}\right)^2 \exp(-l/R_d)\right), & R_d \to \infty \end{cases}$$

Now let us find the interpolation for $\Phi_U$ in the case $\theta \ll 1$ (i.e. $\Delta R \ll R_0$). In this case $q_m \approx \left(\dfrac{\varepsilon-1}{\varepsilon+1}\right)^m \dfrac{1}{m+1}$, $r_m \approx R_0\,\dfrac{m}{m+1}$, so one could easy calculate the sum $\sum_{m=0}^{\infty}\left(\dfrac{\varepsilon-1}{\varepsilon+1}\right)^m \dfrac{1}{m+1} = \dfrac{\varepsilon+1}{\varepsilon-1}\ln\!\left(\dfrac{\varepsilon+1}{2}\right)$ and then obtain [1/1] Pade approximation with 5% accuracy for interaction energy $\Phi_U$ over variable $R_d$, namely:

$$\Phi_U(d,l) = \frac{4\pi(\sigma_S - P_S)}{\varepsilon + 1} \sum_{m=0}^{\infty} q_m R_d \frac{\sqrt{(z_0 - r_m)^2 + d^2} - (z_0 - r_m)}{R_d + 2\sqrt{(z_0 - r_m)^2 + d^2}} \approx$$
$$\approx \frac{4\pi(\sigma_S - P_S)UR_0}{\varepsilon - 1} \ln\left(\frac{1+\varepsilon}{2}\right) R_d \frac{\sqrt{z_0^2 + d^2} - z_0}{R_d + 2\sqrt{z_0^2 + d^2}}$$
(C.3)

The approximation in (C.3) is valid under the condition $\Delta R \ll R_0$ typical for experiments.

2) The depolarization field energy related to the surfaces $z = 0$ and $\Sigma$ has the following form:

$$\Phi_D(d,l) = \Phi_{DV}(d,l) + \Phi_{DS}(d,l)$$
(C.4)

The "bulk" depolarization field energy $\Phi_{DV}(d,l)$ has the form:

$$\Phi_{DV}(d,l) = \int_{\Sigma(z>0)} ds(\mathbf{P}_S \cdot \mathbf{n})\varphi_{DE} - \frac{\varepsilon}{8\pi R_d^2} \int_{z>0} dv\, \varphi_{DE}^2 = \begin{cases} \dfrac{16\pi^2 P_S^2}{3(\varepsilon+1)} \dfrac{d^4}{l}(\ln(2l/d) - 1), & R_d \to \infty \\ \dfrac{4\pi^2 P_S^2}{\varepsilon + 1} d^2 R_d, & R_d \to 0 \end{cases}$$

The [1/1] Pade approximation for the "bulk" depolarization field energy $\Phi_{DV}(d,l)$ over variable $R_d$, acquires the following form:

$$\Phi_{DV}(d,l) \approx \frac{16\pi^2 P_S^2}{3(\varepsilon+1)} \frac{d^4 R_d (\ln(2l/d) - 1)}{l R_d + 4d^2 (\ln(2l/d) - 1)/3}$$
(C.5)

Similarly to (C.5), it is easy to obtain from the Gauss theorem, that the expression for the "butt" depolarization energy $\Phi_{DS}(d,l)$ should be rewritten as:

$$\Phi_{DS}(d,l) = \int_{\Sigma(z>0)} ds(\mathbf{P}_S \cdot \mathbf{n})\varphi_{DS} + \int_{\substack{(x^2+y^2)\leq d^2 \\ z=0}} dxdy\, \varphi_{DS} \frac{\sigma_S - P_S}{2} - \frac{\varepsilon}{8\pi R_d^2} \int_{z>0} dv\left(\varphi_{DS}^2 + 2\varphi_{DS}\varphi_{DE}\right)$$
(C.6)

Using (B.8-9), we estimate the integrals in (C.6) under the conditions $R_d \to 0$ and $R_d \to \infty$.

In the case $R_d \to \infty$ we obtained the approximation:

$$\Phi_{DS} \approx \begin{cases} \dfrac{4\pi^2(\sigma_S - P_S)^2}{3(\varepsilon + 1)} d^3 \left(\dfrac{4}{\pi} - (1 + 2\ln 2)\dfrac{d}{l}\right) + O\left(\dfrac{d}{l}\right)^2 & at \quad \Delta R \gg R_0 \quad or \quad d \geq \varepsilon R_0 \\ 2\pi^2 (\sigma_S - P_S)^2 d^2 \Delta R & at \quad \Delta R \ll R_0 \quad and \quad d \ll \varepsilon R_0 \end{cases}$$
(C.7)

In the opposite case $R_d \to 0$ we obtained the approximation:

$$\Phi_{DS} = \begin{cases} \dfrac{\pi^2 (\sigma_S - P_S)^2}{\varepsilon} d^2 R_d & at \quad \Delta R \gg R_0 \quad or \quad d \geq \varepsilon R_0 \\ \pi^2 (\sigma_S - P_S)^2 d^2 R_d \Delta R \dfrac{2R_d + \varepsilon \Delta R}{(R_d + \varepsilon \Delta R)^2} & at \quad \Delta R \ll R_0 \quad and \quad d \ll \varepsilon R_0 \end{cases}$$
(C.8)

Free energy terms for the anisotropic semiconductor can be obtained similarly from the potentials (A.7), (B.3) and (B.8) after the substitution (A.10).